\documentclass[sigconf]{acmart}
\usepackage{listings}
\usepackage{graphicx}
\usepackage{tabularx}
\usepackage{makecell}
\usepackage{subcaption}

\AtBeginDocument{%
  }

\copyrightyear{2025}
\acmYear{2025}
\setcopyright{rightsretained}
\acmConference[CHI '25]{CHI Conference on Human Factors in Computing Systems}{April 26-May 1, 2025}{Yokohama, Japan}
\acmBooktitle{CHI Conference on Human Factors in Computing Systems (CHI '25), April 26-May 1, 2025, Yokohama, Japan}\acmDOI{10.1145/3706598.3713598}
\acmISBN{979-8-4007-1394-1/25/04}

\begin{document}


\title[The Design Space for Online Restorative Justice Tools]{The Design Space for Online Restorative Justice Tools: A Case Study with ApoloBot}

\author{Bich Ngoc (Rubi) Doan}

\email{ngocdb1609@gmail.com}
\orcid{0009-0006-1767-2585}
\affiliation{%
  \institution{KAIST}
  \city{Daejeon}
  \country{Republic of Korea}
}
\authornote{now at EPFL, Switzerland.}

\author{Joseph Seering}
\email{seering@kaist.ac.kr}
\orcid{0000-0001-7606-4711}
\affiliation{%
  \institution{KAIST}
  \city{Daejeon}
  \country{Republic of Korea}
}

\renewcommand{\shortauthors}{Doan and Seering.}

\newcommand{\todo}[1]{\textcolor{red}{TODO: #1}}
\newcommand{\note}[1]{\textcolor{blue}{Note: #1}}
\newcommand{\revision}[1]{\textcolor{black}{#1}}
\newcommand{\revisionn}[1]{\textcolor{black}{#1}}

\begin{abstract}
   Volunteer moderators use various strategies to address online harms within their communities. Although punitive measures like content removal or account bans are common, recent research has explored the potential for restorative justice as an alternative framework to address the distinct needs of victims, offenders, and community members. In this study, we take steps toward identifying a more concrete design space for restorative justice-oriented tools by developing ApoloBot, a Discord bot designed to facilitate apologies when harm occurs in online communities. We present results from two rounds of interviews: first, with moderators giving feedback about the design of ApoloBot, and second, after a subset of these moderators have deployed ApoloBot in their communities. This study builds on prior work to yield more detailed insights regarding the potential of adopting online restorative justice tools, including opportunities, challenges, and implications for future designs.
\end{abstract}

\ccsdesc[500]{Human-centered computing~Empirical studies in collaborative and social computing}

\keywords{Content moderation, online harassment, alternative justice, Discord}
\received{12 September 2024}
\received[revised]{10 December 2024}
\received[accepted]{16 January 2025}

\maketitle

\section{Introduction}
Online harm remains a pressing issue for content moderation. A survey by Ipsos revealed that nearly 60\% of online users reported experiencing some form of harm, while over 40\% of victims refrained from seeking external help~\cite{Dunn2023}. This highlights the shortcomings of traditional moderation approaches, which typically emphasize punishments such as account sanctions and content removals~\cite{Gillespie2018, Roberts2019}. While these methods may stop the harm as it occurs, they often fail to prevent further recurrence and meet the needs of victims and others involved. This perspective has been widely echoed within the HCI and CSCW community: several studies pointed out how punitive approaches often leave users unsupported, leading to negative emotions~\cite{Jhaver2019b, Ma2021}, difficulty making sense of their penalties~\cite{West2018, Vaccaro2020}, and struggles in reforming behaviors~\cite{Kou2021}. In response to these shortcomings, some researchers have advocated instead for alternative frameworks like restorative justice, which prioritizes mediation and engagement among stakeholders to address harm collectively.

Restorative justice focuses on repairing harm by involving offenders, victims, and sometimes community members in conversations aiming at healing and resolution. Within this process, offenders are supported to come to terms with their wrongdoing while victims can voice their needs and receive closure from the community. This approach has been effective in reducing recidivism and supporting reintegration in real-world contexts~\cite{Ness2016, Wood2016}.

Recent years have seen a growing interest in applying restorative justice methods to address harm in online environments. These methods have been proposed as a potential support mechanism for offenders such as demonetized creators~\cite{Ma2021}, banned game players~\cite{Kou2021}, and cyberbullying attackers~\cite{Aliyu2024}. Victims from various social groups also show a preference for restorative  resolution~\cite{Schoenebeck2021a, Schoenebeck2021b, Schoenebeck2023}. 
However, much of the discussion remains conceptual. While it is evident that neither retributive nor restorative justice is a one-size-fits-all solution, and that they are effective only in specific contexts and value systems, the precise conditions behind their differential effectiveness are still not fully understood. Furthermore, practical implementations of restorative justice in online spaces are still underexplored, and it remains generally unclear how restorative practices can be structured and integrated into community moderation.

To address this gap, our study explores the implementation of restorative justice from a design perspective, specifically through tools designed for online communities. We developed ApoloBot, a Discord bot that helps moderators streamline a subset of restorative justice principles within their communities by facilitating apologies between offenders and victims, serving as a probe to explore the design space for restorative justice tools. Though restorative justice encompasses a variety of methods, apologies are one of the most basic and core elements, representing a reasonable starting point for exploration. By embedding a process for apologizing into a bot~---~one of the platform's most commonly used moderation tools~---~we provide a practical, accessible resource to \revision{examine the introduction of} restorative justice methods into the online space\revision{, along with} their broader applicability and potential impacts. 

Through interviews with 16 Discord moderators and deployment with six of them, we gather insights into how moderators perceive and engage with restorative justice tools like ApoloBot. Extending prior work, we conducted a more in-depth analysis of the opportunities and challenges associated with integrating restorative justice tools into existing moderation mechanisms. We seek to answer the following research questions:
\begin{itemize}
    \item \textbf{RQ1:} \textit{What opportunities do restorative justice tools present for online communities? Specifically, what values and contexts determine their effectiveness?}
    \item \textbf{RQ2:} \textit{What challenges accompany the implementation and usage of restorative justice tools in online communities?}
\end{itemize} 

We conclude by discussing the 
\revision{possible ways to evaluate the tools' expected outcomes, and propose design implications}
to drive the ongoing development of restorative justice frameworks in the online space. 
\section{Related Work}
\subsection{Online Moderation Practices and Technological Tools}
Online harm refers to a wide range of problematic behaviors, including but not limited to hate speech~\cite{Mathew2020}, public shaming~\cite{Basak2016}, doxxing~\cite{Snyder2017}, and digital self-harm~\cite{Pater2017}. These actions occur in various contexts, including self-directed, collective, or interpersonal victim-perpetrator relationships~\cite{Krug2002}. Our research focuses on the latter, specifically interpersonal harm that involves targeted victims harmed by one or more individuals. 

On platforms such as Reddit and Discord, a subset of users volunteer as moderators, who oversee and manage spaces in order to encourage more positive and productive interactions~\cite{Seering2019}. 
These moderators employ various measures to regulate content and mitigate harm, often reflecting their underlying values and moderation styles~\cite{Jiang2023}. Most commonly, reactive measures such as account sanctions (timeouts, bans) and content removal are used~\cite{Srinivasan2019, Gillespie2018, Roberts2019}. These punishments, typically specified in community rules~\cite{Fiesler2018} and policies~\cite{Schaffner2024}, are enforced after harm has happened. While effective at stopping the immediate harm, excessive sanction can deter engagement and alienate members~\cite{Squirrell2019}, or be criticized as a ``black box'' where punishments are opaque and unfair~\cite{Kou2021, Vaccaro2020}, in some cases driving users to leave communities entirely in favor of other platforms~\cite{Gao2024}. To counter these effects, some moderators integrate proactive strategies to prevent harm before it occurs. This includes establishing social norms~\cite{Seering2019, Chandrasekharan2018}, creating content filters~\cite{Jhaver2019}, and highlighting examples of prosocial behaviors~\cite{Seering2017}. 

However, achieving such extensive goals can be time-consuming, labor-intensive~\cite{Steiger2021}, and emotionally draining~---~especially for volunteer moderators who invest significant time and effort into their work~\cite{Wohn2019}, akin to a “second job” for some~\cite{Seering2022a}. To assist moderators, a variety of tools and agents have been developed, including user-developed or third-party bots, algorithms, and artificial intelligence~\cite{Jhaver2019, Kiene2020, Hwang2024}. These tools help in tasks both proactively, such as setting up automatic content filtering~\cite{Chandrasekharan2019}, \revision{sending nudges to potential violations~\cite{Seering2024}}, classification~\cite{Blackwell2017}, and limiting the reach of content~\cite{Binns2017}, and reactively, such as issuing warnings, bans, and handling user appeals~\cite{Atreja2024}. In addition, tools like visual exploration systems can combine both proactive and reactive strategies~\cite{Choi2023}.

In a broader sense, indirect measures such as education~\cite{Cai2019} and community support~\cite{Kou2024} also play a crucial role in addressing harm and reforming users before, during, and after harmful incidents. There has been growing interest in applying notions of justice to online moderation, focusing on user rehabilitation and reparation. 
Researchers argue for a shift from punitive interventions to a greater emphasis on user education and restorative forms of justice-seeking~\cite{West2018, Blackwell2018}. Cai et. al reveals that a caring approach can convert one-time offenders into long-term loyal members~\cite{Cai2021}. Restorative justice, in particular, is gaining traction as a potential approach~\cite{Kou2021} to address harm holistically, taking into account victims’ specific needs while engaging offenders and community members to heal collectively. Under this justice lens, profound studies have explored the central needs of harmed adolescents~\cite{Xiao2022}, in addition to different contextual challenges and opportunities for implementing restorative approaches online~\cite{Xiao2023}. Nevertheless, given that each justice approach has pros and cons, repairing harm is not a one-size-fits-all endeavor~\cite{Schoenebeck2021a, Warzel2019}. Different harms might require distinct frameworks or a combination of different ones~\cite{Goldman2021, Llewellyn1999}. Our research builds on this by investigating the design space with more in-depth contexts where online restorative justice implementation may prove the most effective.

\subsection{Restorative Justice and its Implementation in the Online Space}
\subsubsection{Restorative Justice Overview}
Restorative Justice is a framework that centers on people’s needs by providing care and support after harm occurs. This differs from the common punitive model, where harmful behaviors are considered rule violations that require punishments in proportion to their offense~\cite{Szablowinski2008}. Restorative justice extends beyond this depiction to view harm as “a violation of people and relationships rather than merely a breach of rules”~\cite{Ness2016}. It embodies three main principles~\cite{Mccold2000}: (1) identify and address victims' needs related to the harm; (2) hold offenders accountable to right those wrongs; and (3) involve victims, offenders, and potentially the community in the restoration process. Instead of punishing the offenders, the primary tool for a restorative approach is communication among involved stakeholders, usually mediated by a facilitator~\cite{Bolitho2017} in offline settings. The facilitator helps ensure that victims and offenders have equal footing, guides them to reflect on the harm, and determines whether a consensus can be reached without incurring further harm. This approach has been successfully applied in settings such as the criminal justice system, schools, and workplaces~\cite{Ness2016, Wood2016}.

\subsubsection{A Prevalent Restorative Justice Approach - Apology}
As applied to the online landscape, restorative justice can manifest in many different ways, but one of the most prevalent is through \textbf{apology}. Schoenebeck et al. explored various forms of restorative intervention \revision{to online harassment}, including mediation, identity education, and notably, apology, which was strongly preferred by victims from various social groups, youths, and across countries~\cite{Schoenebeck2021a, Schoenebeck2021b, Schoenebeck2023}. In co-designing to address cyberbullying, \revision{structured restoration systems were proposed where apology served as a condition for resolution, requiring offenders to complete empathy-building training and reconcile with victims before having their punishments lifted~\cite{Aliyu2024}. Applying restorative justice principles, Xiao et al. identified five key needs among harmed adolescents: sensemaking,
emotional support and validation, safety, retribution, and transformation~\cite{Xiao2022}. Apologies specifically addresses two of these needs~---~providing emotional support and validation by prompting offenders to acknowledge their wrongdoing, and enhancing safety by a commitment to cease further harmful behaviors.} Despite not being formally integrated into online governance, volunteer moderators sometimes solicit apologies from offenders, or deliver apologies to targets themselves~\cite{Matias2019, Seering2019}. Given the widespread relevance and importance of apologies in online contexts, we employ this concept as the core framing for our restorative justice tool implementation, 

\subsubsection{Implementing Restorative Justice Online}
Studies have explored various restorative justice implementations in online environments, both manually and technically. As a manual approach, Xiao et. al investigated the viability of the \textit{victim-offender conference}~---~a widely used practice for restorative justice~\cite{Xiao2023}. Under this setup, the moderator, acting as a guiding facilitator, holds separate meetings with the victim and the offender before both parties agree to meet and discuss the harm, potentially reaching a resolution~\cite{Zehr2015}. While promising, particularly for victims, the process is hindered by its labor-intensive nature and the prevailing stigma of punitive systems. Alternatively, technical approaches such as Keeper~\cite{Hughes2020} were introduced, providing an environment to facilitate online restorative justice circles through visual features and spatial attributes. Despite its efficiency in improving interaction quality, Keeper’s separate platform and unfamiliar approach may present adoption and integration problems. Learning a new system demands additional effort and commitment from moderators, especially for content moderation at scale~\cite{Kiene2019}. This is compounded by the fact that many community moderators are unpaid volunteers and already overburdened~\cite{Wohn2019}, adding further pressures in shifting their moderation framework.

These challenges highlight the need for more practical and scalable implementation of restorative justice within online communities.
\revision{
One effective path is to build on toolsets that moderators already rely on, rather than introducing entirely new systems~\cite{Kuo2023}. In many communities, particularly on platforms like Discord, custom bots serve as integral tools to facilitate social process \cite{Hwang2024}, adapt to changing needs~\cite{Kiene2019}, and manage growing membership base~\cite{Kiene2020}. Accordingly, our study builds on this precedent by exploring bots as a familiar, adaptable framework to facilitate stakeholder engagement in a restorative process similar to a victim-offender conference.
}


\subsubsection{Evaluating Technological Restorative Justice Tools}
Assessing the effectiveness of alternative methods for addressing online harm is important, yet challenging due to the complex social factors involved~\cite{Ma2023, Aliyu2024}. This inquiry extends to restorative justice with key questions: Under which scenarios would restorative implementation make a meaningful difference, and when might it not be worth pursuing?~\cite{Cai2024} How can this approach be implemented as a practical tool, and what are the expected results if it proves effective?~\cite{Xiao2023}

The nuances to restoring interpersonal harm, highlighted by Kou et al., underline the importance of considering “contextuality of toxicity” as a factor for designing restorative mechanisms~\cite{Kou2021}. Xiao et al. argue that restorative justice application should be a value-related question, reflecting one’s moderation goals~\cite{Xiao2023}. These insights suggest that the opportunities for restorative justice tools are heavily influenced by contextual factors. Building on these findings, our work examines the opportunity space of restorative tools across three different scopes: community, moderation practice, and case scenario, which are essential areas of focus for tool adoption, implementation, and effective usage, respectively. Utilizing ApoloBot as a conceptual implementation of restorative justice, we gather insights from moderators through interviews and deployments, identifying specific examples and situations where this approach may or may not be useful. We further pinpoint practical challenges associated with the tool implementation and propose implications for future designs.
\section{System Design: ApoloBot}
\subsection{Context: Discord and Its Moderation Practices} \label{context}
In this study, we choose Discord as our research site due to its community-oriented social structure and the flexibility of its API for tool design. Unlike traditional platforms that emphasize individual profiles, Discord is centered around the concept of \textit{servers}, where communities are formed from small groups of friends to large circles with millions of people. Originally created for gamers as a third-party voice function, Discord servers now serve a variety of topics such as technology, art, and entertainment. Servers can be public~---~where a link is posted online for anyone to join~---~or private, with invitations more strictly limited. Within these spaces, members interact with each other in \textit{channels}, either through text or voice chat. Channels are areas in the server serving specific purposes, such as announcements, general chats, and topic-specific discussions. Moderators in these spaces are usually volunteers who are active community members, though those in some more formally structured servers may be paid. While their responsibilities vary from one server to another, the overarching goal is to ensure the community's safety and well-being as it grows. Moderators' duties may involve establishing community guidelines, engaging in conversations, and supporting members facing issues within the server. When incidents occur, moderators can take action using Discord's built-in functions like mute, ban, or message removal. Recently, Discord introduced AutoMod,\footnote{https://discord.com/tags/automod} an automated feature that assists moderators with tasks like content filters, action settings, infraction logging, and user verification. To further streamline and customize moderation efforts, third-party bots are extensively adopted, being used by nearly one-third of all Discord servers~\cite{Warren2021}. These bots perform various functions, including tracking server status, managing member activities, and issuing moderation actions, among many others. Popular examples are MEE6,\footnote{https://mee6.xyz/} Zeppelin,\footnote{https://zeppelin.gg/} Tatsu,\footnote{https://tatsu.gg/} and Dyno.\footnote{https://dyno.gg/} These automated solutions help alleviate the workload on human moderators, enabling them to focus on more significant issues while bots manage more routine tasks. Bots can also bring a sense of fun and engagement with features like welcome messages, role assignments, and customized activities. 

In addition to its community structure, Discord offers extensive APIs and interactive features for creating custom bots. These accessible resources add to our decision to utilize the platform and develop ApoloBot~---~a tool that embeds restorative justice principles into online community moderation. 

\subsection{System Implementation} \label{system} 
\begin{figure*}[h!]
  \centering
  \includegraphics[width=0.9\textwidth]{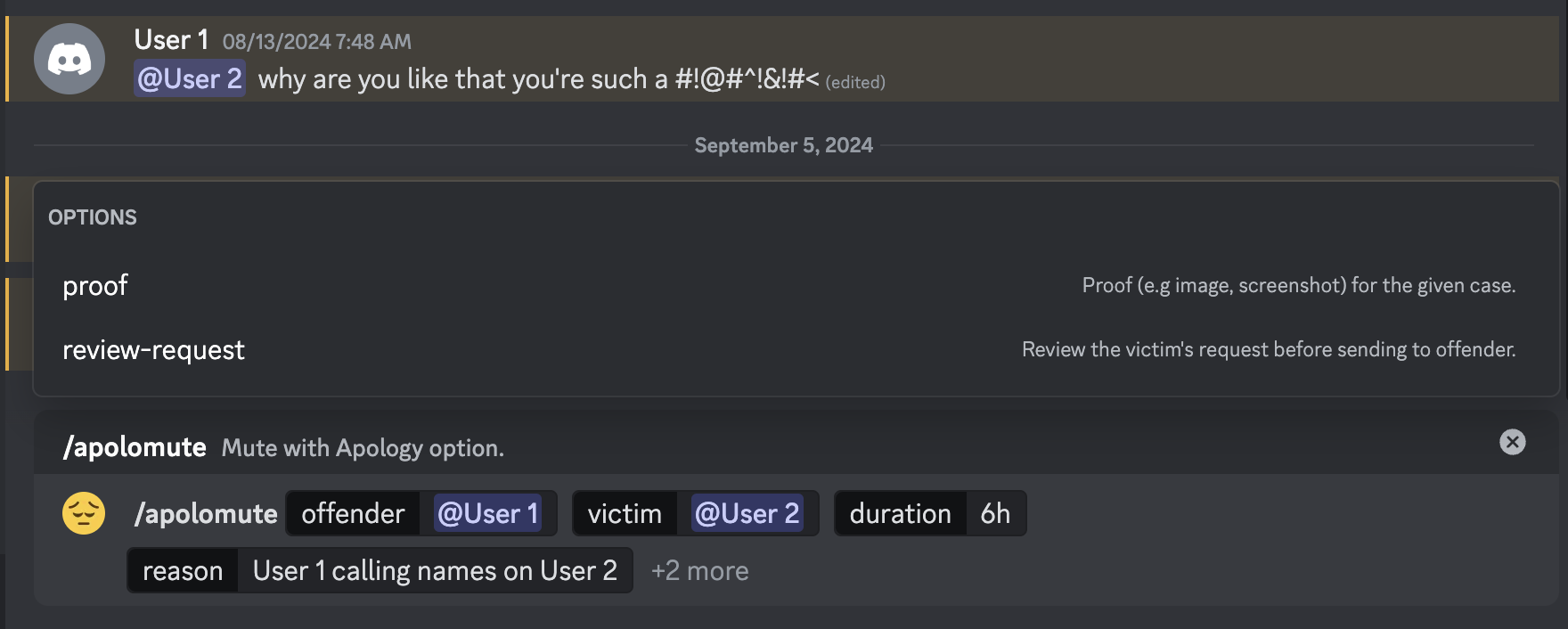}
  \caption{The slash command \textit{/apolomute} that is used in the primary workflow. The first four input fields are required, where the moderator specifies the involved offender and victim, along with mute duration and reason. Optionally, proof can be attached as an image, and moderator can choose to first review the victim's apology request by setting `review-request' to True.}
  \Description{The message at the top shows an example of a harmful interaction between two users, involving some insults. Below, there is a chat box where the slash command is inputted. The command begins with a slash (/), followed by its name (apolomute), and includes several required fields: offender, victim, duration, and reason. It also suggests optional fields such as proof and review-request above the chat box.}
  \label{fig:slashcmd}
\end{figure*}

ApoloBot was developed with Javascript and operates on Node.js. It utilizes MongoDB as its database and Heroku as the hosting service. The discord.js library was employed to access necessary Discord APIs for managing user interactions and bot features. The bot’s core functionality is based on the slash command \textit{/apolomute}, for which syntax is shown in Figure \ref{fig:slashcmd}. Slash commands are familiar formats among Discord moderators, where punishments are typically executed via \textit{/mute} or \textit{/ban} commands provided by Discord's built-in system or other bots. Moderators can choose different slash commands based on the situation, allowing \textit{/apolomute} to work alongside other moderation commands. This flexibility makes ApoloBot easy to learn and integrate into any moderator's existing framework.

\subsection{Workflow} 
ApoloBot's procedure draws inspiration from the concept of the \textit{victim-offender conference}, a model for online restorative justice practices explored in Xiao et al.’s case study~\cite{Xiao2023}. Within this framework, victims and offenders are encouraged to meet, reflect, and resolve the harm under the guidance of a facilitator, who is the moderator in this context. The moderator's role is to ensure that the process remains safe and constructive, with the final decision ideally determined by the victims and offenders. While the original approach is manual, ApoloBot facilitates a version of this process that is tracked and guided in a style familiar to moderators who are experienced with other Discord bots. 

\begin{figure*}[h!]
  \centering
  \includegraphics[width=\linewidth]{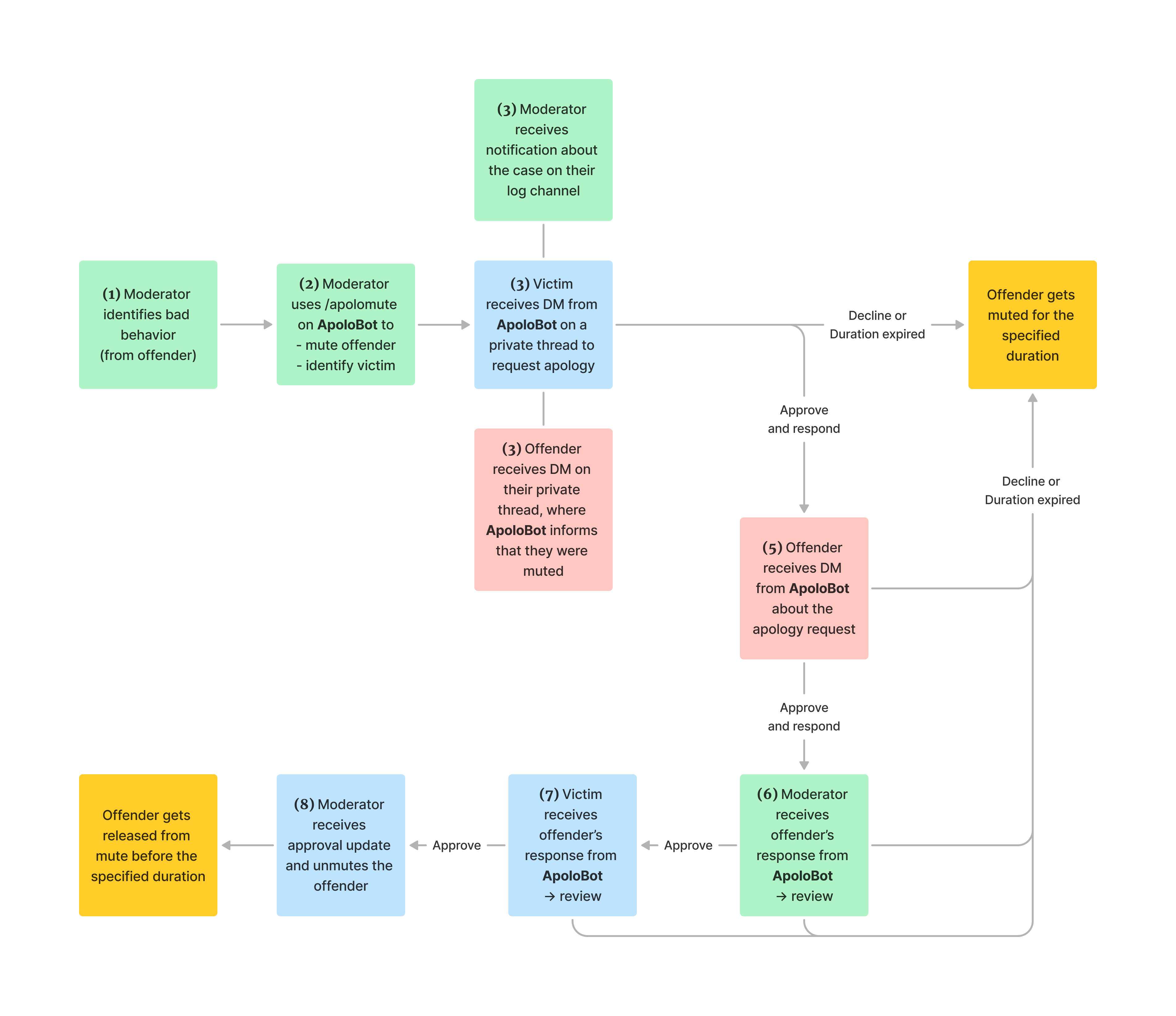}
  \caption{ApoloBot's Primary Workflow. The diagram shows the different pathways ApoloBot can follow based on stakeholders' decisions to approve or decline their actions. The green blocks represent the interaction points for the moderator, who keeps up with ApoloBot through their log channels. The blue and red ones depict the interaction for the victim and the offender, respectively, in their private threads. Yellow blocks indicate the case is closed and no further steps will be taken.}
  \Description{The start block is in the top left, where "Moderator identifies bad behavior (from offender)" and the workflow is initiated. From here, the process might follow different pathways depending on stakeholders' participation. If everyone approves and responds to ApoloBot prompts, they will reach the end block in the bottom left where "Offender gets released from mute before the specified duration". Otherwise, they end up with the block in the top right, which indicates "Offender gets muted for the specified duration".}
  \label{fig:workflow}
\end{figure*}

\revision{
Commonly, commands like \textit{/mute} or \textit{/ban} are utilized as a standardized procedure to impose punishment terms on the offender following an incident of harm. However, this approach might perpetuate a punitive mindset that deters meaningful engagement when switching to restorative justice~\cite{Xiao2023}. Gradual changes are typical in real-world justice systems, and sometimes a dual system incorporating both punitive and restorative measures is employed to cater to the diverse stakeholder needs while adapting to established structures
~\cite{Llewellyn1999}. Building on this, ApoloBot's \textit{/apolomute} extends the conventional mute by offering a potential restorative interaction point, opening an avenue for apology and constructive resolution while retaining certain familiarity with the mute action.} Figure \ref{fig:workflow} outlines an overview of ApoloBot's primary workflow, incorporating the \textit{/apolomute} slash command.

The workflow begins similarly to \revision{many current moderator response flows:} upon recognizing inappropriate behavior, the moderator mutes the offender. However, ApoloBot adds a step by involving the victim to initiate the apology process with the offender, and moderators facilitate this by specifying both the victim and the offender in the slash command syntax (Fig \ref{fig:slashcmd}). Once the command is executed, ApoloBot creates two separate threads\footnote{https://support.discord.com/hc/en-us/articles/4403205878423-Threads-FAQ}~---~one for the victim and one for the offender~---~where the interaction between ApoloBot and these stakeholders will take place. For the moderators, they interface with ApoloBot through a dedicated log channel.

\begin{figure*}[btp]
  \centering
     \begin{subfigure}{\textwidth}
         \centering
         \includegraphics[width=\textwidth]{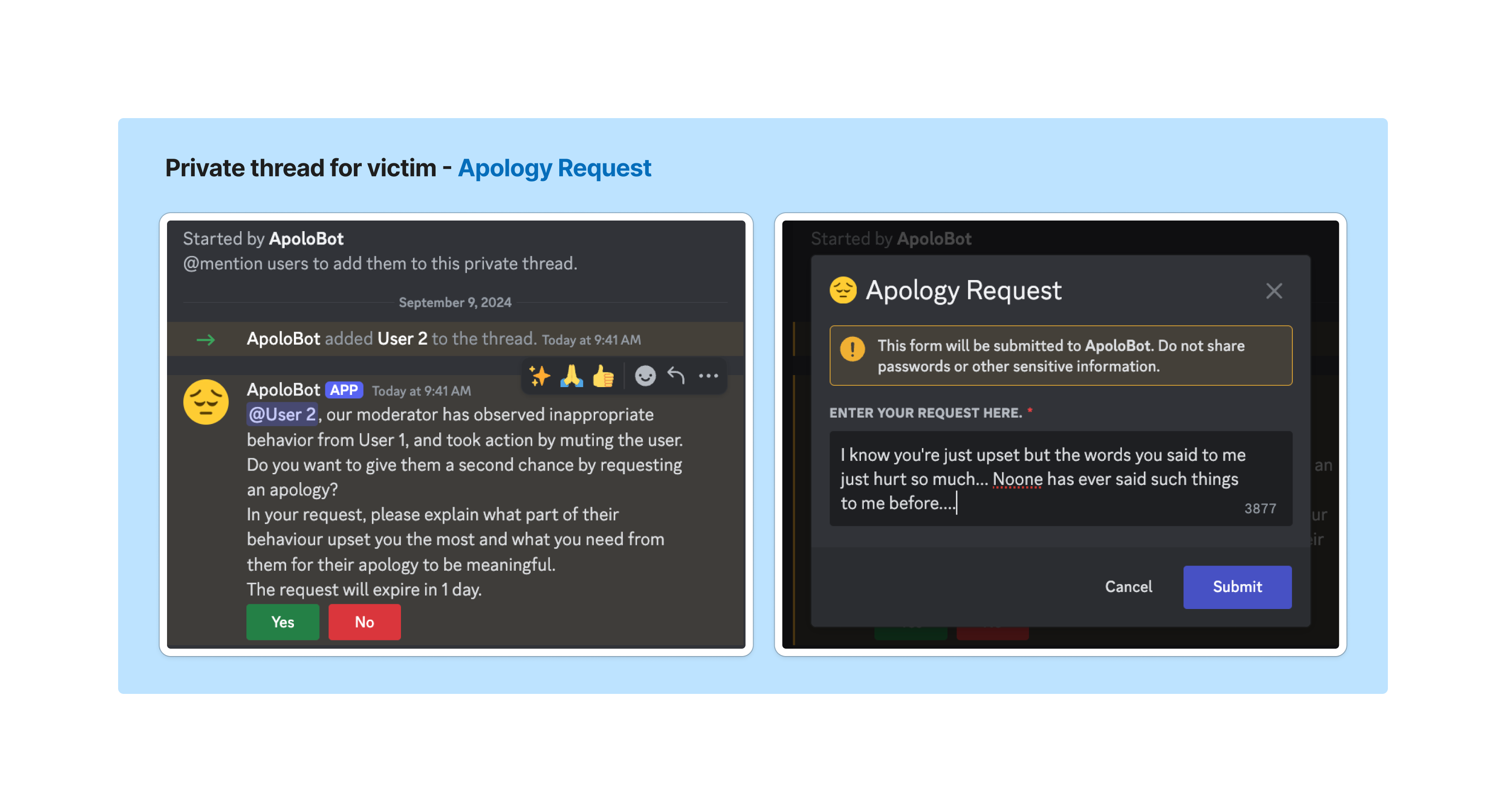}
         \caption{Victim's private thread for requesting an apology}
         \label{fig:private-thread-victim}
         \Description{The two pictures illustrate the interaction between ApoloBot and the victim in their private thread. The first picture shows the text message ApoloBot sends, which informs the victim about the case and asks if they want to request an apology. Two buttons below are provided for the victim's decision: "Yes" and "No". The second picture shows the subsequent screen after the victim selects "Yes". ApoloBot displays a popup text box, with a text field for the victim to fill in their apology request. In the bottom right, there is a "Submit" button to send this request.}
     \end{subfigure}
     \hfill
     \begin{subfigure}{\textwidth}
         \centering
         \includegraphics[width=\textwidth]{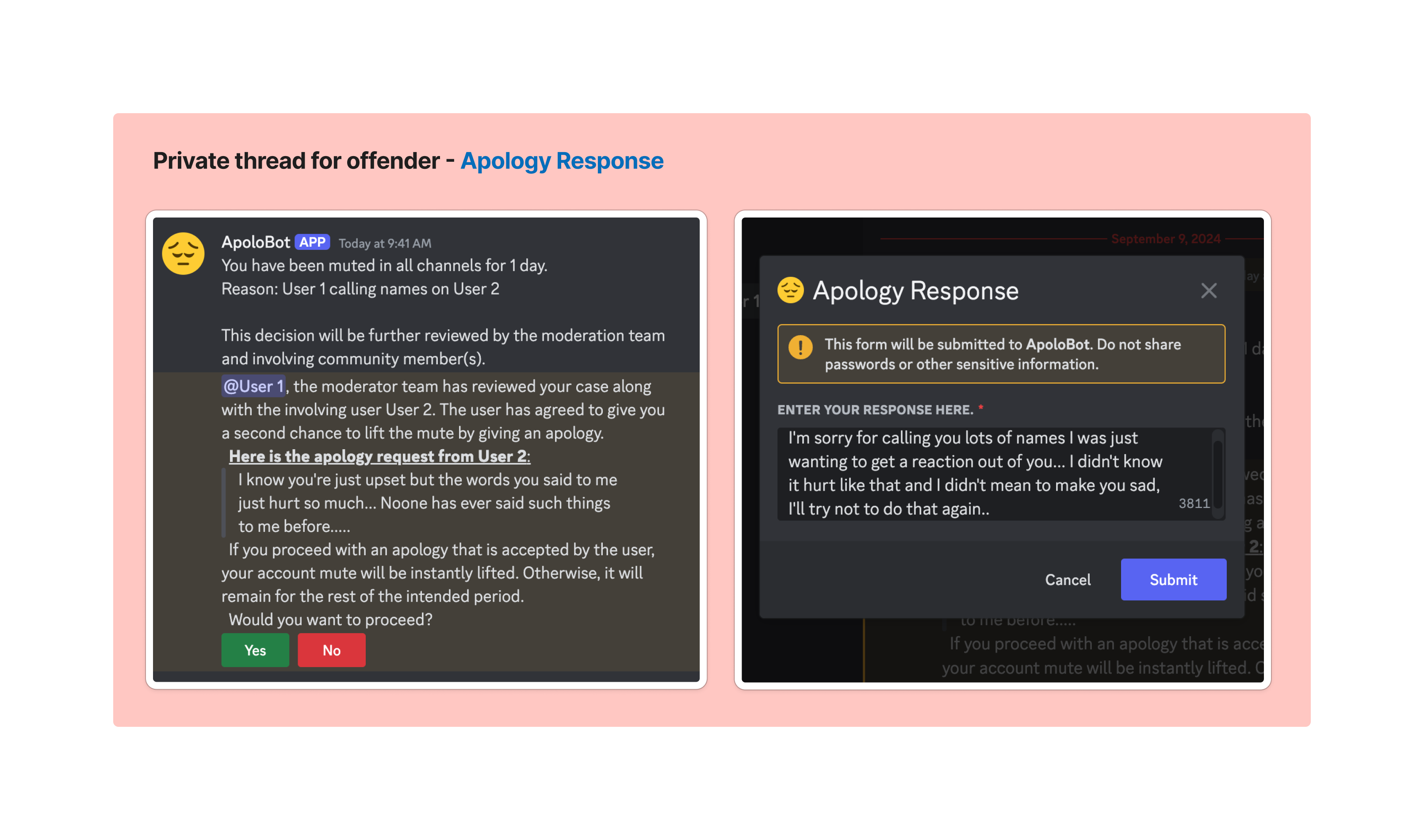}
         \caption{Offender's private thread for requesting an apology.}
         \label{fig:private-thread-offender}
         \Description{The two pictures illustrate the interaction between ApoloBot and the offender in their private thread. Similarly, the first picture shows ApoloBot's text message. It contains the quoted apology request from the victim, and asks if the offender wants to respond with an apology. This is followed by two "Yes" and "No" decision buttons. The second picture also shows the next screen after selecting "Yes", where ApoloBot shows a popup text box. The offender fills in their apology response in a text field, and sends this via the "Submit" button.}
     \end{subfigure}
  \caption{Examples of private threads ApoloBot created from the victim's side and the offender's side.}
  \Description{}
  \label{fig:private-threads}
\end{figure*}

In the victim's private thread, ApoloBot informs them that the offender has been muted for harmful behavior and offers the option to request an apology. If chosen, this option grants the offender a second chance to make amends and potentially have the punishment lifted. If the victim chooses to proceed, they are prompted to enter their apology request via a popup textbox (Fig \ref{fig:private-thread-victim}). 

Following this, ApoloBot notifies the offender in their private thread that their mute can be lifted if they deliver an appropriate apology to the victim. If the offender decides to comply, they are prompted to write their apology response via a similar popup textbox (Fig \ref{fig:private-thread-offender}). 

\begin{figure*}[h!]
  \centering
  \includegraphics[width=0.6\textwidth]{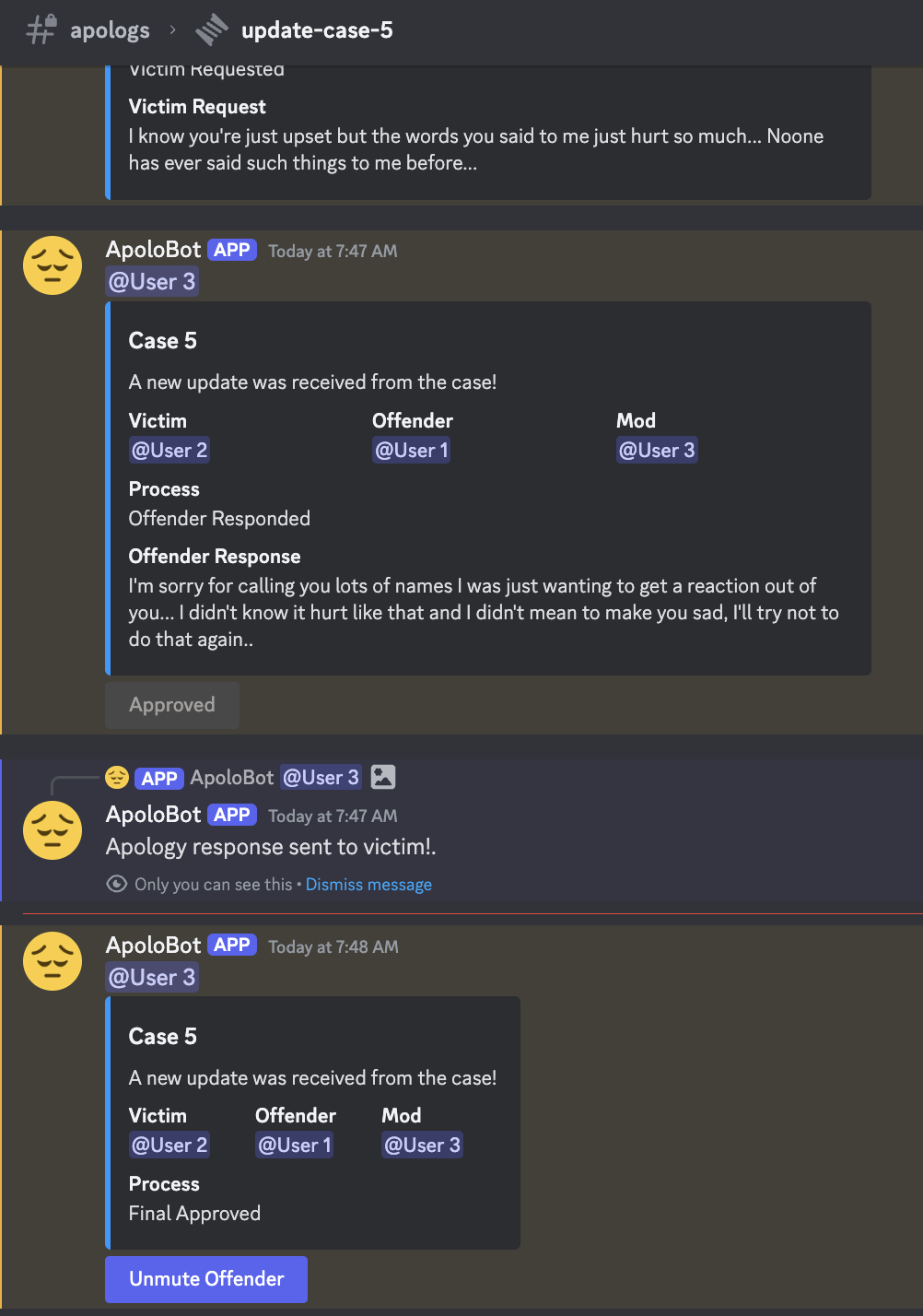}
  \caption{Examples of ApoloBot logs received by moderators.}
  \Description{The picture shows a series of logging messages sent by ApoloBot to the moderator, providing updates on a specific case. On the top bar, the name of the log channel is shown, followed by the thread name (update-case-5). The main interface presents messages in distinct blocks, each detailing information such as the victim and the offender's name, the current process step, and specific updates like the Victim Request or Offender Response. In the final (bottom) message block, it indicates that everyone in the process has approved, and includes a button labeled "Unmute Offender" that the moderator can click to perform the action.}
  \label{fig:logs}
\end{figure*}

Throughout the process, moderators receive updates from ApoloBot at every step. After receiving the apology request and response, they are responsible for reviewing the offender’s response to ensure its appropriateness (Fig \ref{fig:logs}). If approved, the apology is forwarded to the victim, who then has the final say. If the victim accepts the apology, ApoloBot notifies moderators and they can unmute the offender accordingly. 

This approach translates the traditional \textit{victim-offender conference} into a technical process via ApoloBot: The interactive exchange between offenders and victims allows each party to voice their perspectives when deciding the restoration outcome, and the facilitation of moderators ensures this process goes smoothly without incurring additional harm. 

At any stage, if the victim, the offender, or the moderator declines to proceed, or if the designated time expires, the process reverts to the standard punitive measures where the offender remains muted for the specified duration. This is in line with real-life restorative practices, where complete consensus may not always be feasible. 
\revision{Forcing forgiveness from victims or remorse from offenders, however, may compromise the victims' autonomy and lead to disingenuous offender responses~\cite{Llewellyn1999, Bazemore2015}.}
The system therefore supports partial participation, ensuring that engagement is voluntary and all stakeholders’ decisions are respected.
\section{Methods}
Using ApoloBot as a discussion starting point, we extend our exploration into the broader landscape of restorative justice tools through a three-phase user study with Discord moderators. Each phase involves increasing levels of commitment, starting with initial interviews, followed by tool deployment, and concluding with reflections. Given that restorative justice tools are still relatively rare in online communities, these separate phases allow us to gather valuable insights while respecting moderators' diverse willingness and interest in the new approach. All parts of this study were pre-approved by our university's Institutional Review Board (IRB).

\subsection{Phases Overview}

\textbf{Phase 1. Onboarding Session (60-90 minutes):} In the first phase, we conducted individual interviews with Discord moderators to gain insights into their general moderation practices and the potential of integrating restorative justice tools. Participants were asked about their procedure to handling interpersonal harm with specific examples of past cases. We then introduced the concept of restorative justice and presented ApoloBot as a practical tool embodying a subset of these principles. This was followed by discussions on the potential application of ApoloBot and other restorative justice tools within their communities, considering critical factors such as use cases, challenges, opportunities, perceived benefits, and drawbacks. After the interview, participants were invited to a Discord sandbox server to test out ApoloBot, where they provided further feedback and decided whether to continue with the study by deploying it in the subsequent phase.

\textbf{Phase 2. In-the-wild Deployment (4 weeks):} In the second phase, a subset of interested participants deployed ApoloBot in their communities, using it whenever suitable cases arose. Throughout this period, they kept track of their bot usage and maintained weekly communication with the researchers for feedback and support.

\textbf{Phase 3. Exit Interview (60-90 minutes):} At the end of the deployment period, participants joined an exit interview to reflect on their experiences with ApoloBot, unveiling new insights into its practical aspects, including user engagement and its effects on the community. Building on these reflections and revisiting critical factors from Phase 1 interviews, we \revision{explored the underlying factors for how the deployment met or challenged initial expectations, and} broadened the discussion to assess the overall design space of ApoloBot and other online restorative justice tools.

All interviews were conducted remotely through the Discord voice chat function. Participants could withdraw from the study at any phase without penalty. Compensation was provided for fully completed phases: \$20 for Phase 1, \$50 for Phase 2, and \$30 for Phase 3, delivered via Tremendous.~\footnote{https://www.tremendous.com/}

\subsection{Recruitment and Selection of Participants}
Our recruitment call was distributed in meta-moderation communities on Discord, Reddit, and Facebook. These are communities where Discord moderators gather to discuss various moderation topics, such as news, strategies, philosophies, and tool usage. To ensure the quality of our recruits, we used a screening survey to assess their background and moderation experience. In addition to project-specific criteria such as prior experience handling interpersonal harm and familiarity with Discord bots, we filtered out low-quality responses such as one-word answers and those containing nonsensical or irrelevant information. We contacted selected participants, and further employed snowball sampling~\cite{Biernacki1981} by asking them for referrals. A total of 16 participants were chosen for Phase 1, with six proceeding to Phases 2 and 3. Two used ApoloBot during their deployment, while the others deployed it but did not encounter any suitable use cases. A summary of the participants' demographics and their status within Phase 1 and 2 are detailed in Table \ref{table:demographics}.
\newcolumntype{Y}{>{\centering\arraybackslash}X}
\newcolumntype{Z}{>{\centering\arraybackslash}m{4em}}

\begin{table*}[h!]
\begin{tabularx}{\linewidth}{ |Z|Y|Y|Y|Y| } 
\hline
\textbf{P\#} & \textbf{Server Category}       & \textbf{Server Size}     & \makecell{\textbf{Continued after} \\ \textbf{Phase 1?}} & \makecell{\textbf{ApoloBot was used} \\ \textbf{in Phase 2?}} \\
\hline
P1                 & Fandom                         & 5,001 - 10,000 members  & No                                & -                                      \\
P2                 & Content Creator                & Over 10,000 members & No                                & -                                      \\
P3                 & Content Creator, Lifestyle     & Over 10,000 members & Yes                               & No                                     \\
P4                 & Fandom, Gaming                 & Over 10,000 members & No                                & -                                      \\
P5                 & Fandom                         & Over 10,000 members & No                                & -                                      \\
P6                 & Gaming                         & 1,001 - 5,000 members   & Yes                               & No                                     \\
P7                 & Art, Content Creator           & 5,001 - 10,000 members  & Yes                               & Yes                                    \\
P8                 & Content Creator                & Over 10,000 members & Yes                               & No                                     \\
P9                 & Gaming                         & Over 10,000 members & No                                & -                                      \\
P10                & Gaming                         & Over 10,000 members & No                                & -                                      \\
P11                & Gaming                         & 100 - 500 members       & Yes                               & Yes                                    \\
P12                & Content Creator, Mental Health & 1,001 - 5,000 members   & No                                & -                                      \\
P13                & Content Creator, Gaming        & Over 10,000 members & No                                & -                                      \\
P14                & Gaming                         & 1,001 - 5,000 members   & No                                & -                                      \\
P15                & Fandom, Gaming                 & 5,001 - 10,000 members  & No                                & -                                      \\
P16                & Collaboration Platform         & 1,001 - 5,000 members   & Yes                               & No                                     \\                                 
\hline
\end{tabularx}
\caption{Demographics of moderator participants who participated in the user study. \textit{Server Category} refers to the broad topic of the community, for instance Content Creator can include Youtuber or Streamers, and Gaming can include competitive games such as League of Legends and casual plays like Party Animals. Some servers spanned multiple categories, such as a streamer playing specific games (Content Creator, Gaming), or a game with distinct characters and lore (Fandom, Gaming). While most moderators managed multiple communities, we mainly focused on their primary server or the one they deemed most relevant to the study.}
\label{table:demographics}
\end{table*}

\subsection{Qualitative Analysis}
Interview sessions in Phase 1 and Phase 3 were transcribed using CLOVA Notes. 
\revision{
For Phase 1 interviews, thematic analysis was conducted inductively through multiple iterations~\cite{Braun2006}. First, two researchers individually performed line-by-line open coding on eight interviews, generating initial codes that closely resembled text from the transcript such as “instant ban”, “add to modmail” and “bot seems insincere”. This was followed by focused coding~\cite{Saldaa2021}, where we identified recurring themes and sorted them into broader categories such as “escalating procedure”, “integration into existing system” and “tool perception”, which formed our initial codebook. The first author then applied this codebook across the remaining interviews, refining and adding codes as new insights emerged. After this, the two researchers met again to validate the updated codebook, consolidating higher-level themes along the dimension of moderators’ practices in addressing interpersonal harm, their stances on adopting restorative justice tools through ApoloBot’s framework, and potential impacts of implementing such a system. Finally, we aligned these diverse perspectives to outline the opportunity space and challenges associated with transitioning from traditional moderation practice to integrating restorative justice tools, laying the groundwork for our results.
}


Phase 3 interviews were coded by the first author following a similar inductive process \revision{based on the codebook developed in Phase 1. While Phase 1 interviews focused on moderators’ reflections on prior experiences, Phase 3 expanded upon these by grounding the insights in practical deployment outcomes. Successful use cases from Phase 3 demonstrated how expectations from Phase 1 were met, validating the key opportunities where the tool effectively fulfilled its design intent. Equally significant were the unmet expectations, where anticipated use cases were not realized, as they revealed a new-found understanding of the practical challenges and critical areas of the opportunity space where the tool's effectiveness fell short. These observations were thus incorporated into the final codebook by combining and adding to Phase 1's codes, enhancing the framework that underpins our findings.
}


\subsection{Methodological Limitations}
As highlighted by Xiao et al., \textit{"Online communities should allow for partial success or no success without enforcing the ideal outcome, especially at the early stage of implementation when there are insufficient resources or commitments"}~\cite{Xiao2023}. Restorative justice, being relatively new and context-specific, poses significant challenges when evaluated within a brief testing period. Our study is therefore constrained by the limited empirical data available on ApoloBot usage, and the analysis presented here relies mostly on interview data from Phases 1 and 3. 
\revision{
This limitation also arises from how we shape our research focus, which is not on delivering a fully-realized restorative justice tool ready for adoption, but on developing a conceptual artifact to probe its implementation and foster critical reflections among moderators. For those who engaged with the tool, their experiences provide concrete evidence of its realized potential for effective adoption. On the other hand, investigating those who did not use the tool reveals challenges and critical gaps in its suitability within the broader online landscape, which can inform future alternatives or complements that might address the limitations.
Centering the discussion on these dual perspectives allows a deeper and more comprehensive view of how diverse online communities are currently positioned for restorative justice tools, however it might compensate the technical significance of the proposed system.}


In addition, our study primarily gathers insights from moderators rather than victims or offenders. While this focus offers rich insights into the practical aspects of the tool adoption and execution, it lacks the perspectives of the remaining stakeholders essential to restorative justice, and thus may not fully capture the complete user experience.

Finally, even though our participants come from a wide range of international communities covering diverse topics, the fact that they are solely English speakers limits the cross-cultural generalizations that can be made based on our findings.

\section{Findings: The Opportunity Space for Online Restorative Justice Tools} \label{findings1}
Using ApoloBot as an example and starting point for discussing the broader landscape of online restorative justice tools, we present findings in relation to the opportunities and challenges highlighted by our research questions. First, this section addresses the potential for restorative justice tools \textbf{(RQ1)} by examining the opportunity space--a framework that outlines the conditions and environments where restorative justice tools can be applied effectively and bring positive impacts. Section \ref{findings2} then explores the challenges \textbf{(RQ2)} that accompany the integration of such tools.

We begin by analyzing the opportunity space, which encompasses three main scopes emerging from our interviews, with each serving distinct purposes: \textit{Community}, which influences initial adoption based on the tools' alignment with the community’s values; \textit{Moderation Practices}, determining how well they integrate into existing moderation workflows; and \textit{Case Scenarios}, focusing on the specific situations where they can be most effectively applied.

\subsection{Community: Considerations for Initial Adoption}
The decision to adopt restorative justice tools depends on specific community dynamics and characteristics. This section elaborates on the factors influencing communities' readiness for these tools, highlighting the conditions under which initial adoption is most feasible.

\subsubsection{Community Topics and Culture}
The perceived identity of a community is inseparable from the norms for behavior therein, guiding moderation practices. One recurring theme from participant interviews was the influence of a server’s primary focus or topic on the suitability of ApoloBot adoption. Per participants' responses, communities that are most likely to benefit from a restorative approach are ones with a more social focus or themed around human-centric topics. These include servers built around influential individuals or entities, such as content creators, companies, or collaborative platforms. In these spaces, tools are widely adopted for effective organization, and interaction quality among members is also highly valued as it directly impacts the reputation of the central figure or organization. P7, a content creator who moderates their own community, reflected: 
\begin{quote}

\textit{“I have my personal server that's based around the content I make. Here, me and the moderators try to make sure that we have a united fan base, since the more everyone gets along, the better it is for not only the community but also me as the creator, since I can have healthy and engaging audiences. [...] I noticed apologizing allows for really good reconciliation, so it was nice seeing this kind of system being able to make everyone happy in the end. Or at least in good standing with each other.''}

\end{quote}

Healthy resolutions facilitated by a restorative approach not only help the involved parties reconcile but also reflect positively on the influencer. P3 echoes this perspective: \textit{“[It will] align better with their brand, and it would spread, and people would talk about it, and that reinforces good brand equity.''} 

Similarly, moderators of communities that share common interests in topics valuing human interaction and inclusivity~---~such as language learning, mental health, or arts~---~see tools like ApoloBot as valuable for fostering their growth. P12 described these spaces as \textit{“where people are already intended to talk to each other and try to improve as a person''}. Restoration through apologies facilitates deep layers of communication and empathy, supporting these servers' vision of bettering members' well-being.

Conversely, communities that do not prioritize social interaction may find tools like ApoloBot less relevant. For example, servers dedicated to formal or ephemeral topics, like technical support or quick Q\&A exchange, lacked the sustained engagement and relational depth that restorative tools generally require. P11 provided the examples, stating \textit{“I don't think the idea of this bot helps since people are just in the server to solve a specific problem, or get an answer for a specific question, then just leave.”} In communities like this, members have little to no emotional investment in the server or toward each other: \textit{``People will be like `Why should I even apologize'?''} (P3). On the other hand, communities with large volumes of lower-quality social interaction~---~possibly due to negative norms associated with the genre~---~may also struggle to adopt this type of tool. P4, who moderates a server based on a major competitive game, expressed their frustration:

\begin{quote} \textit{“A lot of people get influenced by social media, and they have this mindset that it’s the norm to be toxic here. It’s very extreme, like if you imagine video games are commonly toxic then this game is ten times that. So they would come in the server with this idea in mind and be toxic, and of course they wouldn't even care to apologize even having the option to do so.”} 
\end{quote}

This highlights a fundamental challenge to the adoption of restorative tools; on platforms where inter-community mobility is high and commitment to each member is low, individuals may not be motivated to resolve conflicts constructively. P14 faced a similar issue moderating another gaming server with predominantly young members, describing that the combination of the game's nature and the immaturity of its players lead to people being \textit{``rebellious''} and \textit{``never feel[ing] remorse about what they do.''} 

In these servers, the high incidence of harm might introduce more opportunities for restorative justice tools to be utilized, yet the expected effectiveness might be limited. P14 explains the situation as having \textit{“a greater risk of people abusing the system, but a greater reward of people actually learning what they're doing wrong.”} In fact, many moderators weighed the ``risks'' more heavily than the ``rewards,'' which was a key factor in their decision not to proceed with deploying ApoloBot in Phase 2. In extreme cases, moderators pointed out that some members might not take it seriously, viewing it as a target for jokes or dismissing apologies as \textit{``weak''} or \textit{``cringe''}, further perpetrating the existing negative dynamics.

\subsubsection{Server Size}
The size of the server may also influence tool adoption: Mid-size servers with steady influx of new members are perceived to be the most ideal. They offer spaces where \textit{“there's enough room for people to disagree and potentially cause harm, and for moderators to adequately handle it''} (P3). In contrast, very large servers may experience overwhelming moderation workloads with fewer social commitments: \textit{``The chat goes 300 miles a second. And we look at something, [take] action, then move on with other stuff.'' }(P4) On the other hand, very small servers simply may not find the need for new tools, since moderators may have the capacity to facilitate conversations without needing technical support.


\subsection{Moderation Practice: Fitting into Current Strategies}
Moderators uphold community values through a variety of moderation strategies, which influence whether tools like ApoloBot can be compatible with their existing practices. This section explores how different approaches to moderation~---~viewed as trade-offs rather than strictly positive or negative~---~affect the practical implementation of restorative justice.

\subsubsection{Mediation Approach: Conversation vs. Action}

One critical perspective involves whether moderators' focus aligns with restorative justice values, or whether constructive conversations or punitive actions are prioritized. Successful integration of restorative justice tools requires more than preference; it involves adapting current practices to accommodate the contextual and emotional processes inherent in restorative justice principles. When discussing ApoloBot, moderators pointed out that it is \textit{``more than a binary decision of yes and no, accept or decline''} (P1), as a sense of sensibility is needed to discern first when to use the tool, and then how to evaluate apologies reasonably. This ability is more likely to be found in moderation teams that have established procedures and skills focused on fostering conversations. For instance, P3's workflow features a ``ladder of escalation'' including a defined set of steps that already align fairly well with principles in restorative justice: Interaction, Education, Action, and Moderation. ApoloBot was perceived to fit well within the first two steps where moderators engage with offenders before escalating to further punishments. By contrast, teams with reactive approaches to moderation may lack these perspectives, potentially leading to inefficient tool use and inadequate engagement from members. While these skills can be developed over time, doing so demands significant mental and physical labor, raising barriers for community moderators who are already stretched thin.

Perhaps surprisingly, our user study revealed that the tool appeals most to moderators who align with its values partially but not completely. Those who highly value interaction often prefer to handle conflicts manually, as P9 noted \textit{``Bots should only be assistive, and the key moderation "tool" should be how you portray yourself, and how you listen to other people''}. Conversely, some action-oriented moderators found potential in the tool. P6, whose moderation team previously experienced burnout from high interaction demands to a small moderation scale, sees ApoloBot as a way to ease this burden and re-engage in meaningful conversations with members.

\subsubsection{Flexibility: Fluid vs. Rule-based}

Moderation flexibility determines adaptability. Although apologies are familiar concepts and generally perceived positively, using them as structured tools to resolve online conflicts is not common. This novelty can be viewed in different ways~---~as an opportunity or as a challenge. For moderation teams with fluid dynamics, where procedures are more casual and less rigid, moderators can readily adjust their procedures to experiment with new approaches. Tools like ApoloBot provide offenders a chance to reconcile with victims, but their use is context-dependent and moderators must determine when to offer this leniency. Flexible teams provide time and space for restorative sessions to evolve, improving as moderators gain experience and trust in the process.

On the other hand, systematic and rule-based moderation structures~---~often found in very large or professional servers~---~may find the addition of tools like ApoloBot burdensome or disruptive. Rules on these servers are usually standardized, \textit{“It’s like black and white. You did this so you receive this, you follow the rules or you don't. There is less space to fit restorative justice in between”} (P3). P4 expressed how handling a large-scale server raises the bar for tool adoption and rule enforcement: 
\begin{quote} \textit{``In these places, being "flexible" might mean being messy since the moderation scale is just too large. Consistency and convenience are therefore things we value the most. ApoloBot poses a problem to both of these because firstly, different moderators might evaluate apologies differently, and secondly, the tool would require its own, independent category, which only certain dedicated moderators can handle. Given we already have so many other things going on with hundreds of channels and millions of members, I don't think this addition would be practical.''}
\end{quote}
Deviation from established norms in these servers requires significantly more commitment and resources, therefore moderators in these spaces are less likely to be enthusiastic about adopting tools like ApoloBot.

\subsubsection{Temporal Perspective: Long-term vs. Short-term Goals}
Finally, the moderators' temporal vision defines whether tools like ApoloBot are regarded as ``efficient'' for their moderation practice. ApoloBot is seen to be not favorably “efficient” in the immediate sense, as it takes time to mediate and evaluate ongoing communications among offenders and victims. Some moderators prefer prompt action, especially in highly active communities where interactions, including harmful ones, progress rapidly. \textit{“I’m just looking through the commands channel real quick and see, three days ago there were some guys spamming and being just weird like [sending] NSFW and the n-word. About three people within the span of not even one hour. Then we banned this guy for racism. A day after someone got warned for baiting, someone got banned for a DM spam. [...] People just do the craziest things, so we have to act fast.”} (P6) In these spaces, the high volume and limited time for action demands in-the-moment responses, reducing opportunities for facilitated discussions. On the other hand, some moderators in a more laid-back environment where immediate interventions are less critical, see the long-term ``efficiency'' in a comprehensive and educative approach. As P7 noted:

\begin{quote}
\textit{``Though it takes some time in the time being, making sure everyone gets along in the long run could potentially be more efficient. Because when you're punishing someone and you don't really care too much about apology, people still hold grudges and that could give a lot of drama and bad actors who can upset even more people. And that can create more moderation cases that could create more work for moderators. So it is likely, that by using this system you are removing a lot of future issues that could possibly happen.''}
\end{quote}

In this sense, moderators might spend additional time upfront, if feasible, which can potentially reduce recidivism and thus ease future moderation workload. This long-term investment allows time for tools like ApoloBot to effectively educate users, ultimately fostering more supportive and resilient communities.

\subsection{Case Scenarios: Conditions for Effective Usage}
Once adopted, restorative justice tools must be applied in the right contexts to maximize their impact. This section examines factors specific to the circumstances under which they can be used to resolve conflicts most effectively.

\subsubsection{Types of Interpersonal Harm} Moderators believe emotional and relational harms~---~such as jokes unintentionally coming off offensive, or criticism that turns into insults~---~are often seen as better intended and less serious, thus more fitting for applying restorative justice. These types of harm generally allow for some autonomy in decision-making, and tools like ApoloBot can provide a safe space for people to discuss their issues away from where the harm occurred, thereby helping to defuse emotions and prevent further rages. On the other hand, physical threats (e.g. doxxing) or financial issues (e.g. scams) are considered more severe harms, requiring more immediate and direct intervention. In these cases, initiating conversation might not be appropriate, as it's unlikely to adequately ``pay back'' the caused damage and could even exacerbate the situation. Moderators note that extreme cases may even require higher authorities, such as Discord support or law enforcement, to step in rather than relying on tools.

\subsubsection{Social Ties among Stakeholders} The effectiveness of restorative justice tools is notably enhanced when social ties are present but not overly strong, such as among new members. In these settings, \textit{``the bond is there to appreciate the restoration yet not too much to go out of their way apologizing for the action.''} (P3). Tools can help facilitate these interactions in a less confrontational manner, potentially repairing and strengthening relationships. P11, who successfully employed ApoloBot in several cases, reflected on this impact:

\begin{quote}
    \textit{``Normally, after someone offended others and their mute expired, we keep an eye on how they interact with people, especially the individual they harmed. So I'm seeing in cases I used ApoloBot, the interactions are different if you apologize versus if you don’t. With normal mutes, how these two individuals react after the incident is that they usually become non-friendly, or they hold their bad emotions since the offender wasn’t involved and encouraged to seek reconciliation. With the bot however, after some people had conflicts and they apologized, you could see them becoming normal to each other again since they got all the emotions out. It's something that made me a little bit happy, seeing people react positively afterward.''}
\end{quote}

P11's social game server, where members actively discuss gameplay though aren't closely connected, greatly benefited from ApoloBot in mending relationships after conflicts. In contrast, the tool was perceived as less beneficial at the extremes: With very weak social ties, both offenders and victims may be less concerned about their engagement due to the temporal and anonymous nature of online interactions, and with very strong ties, the involved users are either close friends who would resolve issues privately or users with ``bad blood'' who might refuse to communicate.

Overall, the opportunity space for restorative justice tools can be understood through three key perspectives: \textit{Where} (in which types of community), \textit{How} (through which moderation practices), and \textit{When} (under which scenarios) they are likely to be more or less beneficial. These factors shape the conditions for effective adoption, practice, and usage, as well as the anticipated outcomes and community reactions.

\section{Findings: Challenges of Integrating Online Restorative Justice Tools} \label{findings2}
While restorative justice tools may create opportunities within the outlined space, this doesn't come without drawbacks. Our second research question focuses on the challenges of embedding restorative justice principles within a technological framework, 
\revisionn{
informed primarily by participants’ broader discussions during Phase 1 and also by reflections on ApoloBot’s deployment during Phase 3.
}

\subsection{Adapting to the complex and unpredictable nature of interpersonal harm} 
\subsubsection{Contextual Awareness}
One fundamental challenge for restorative justice tools is capturing the contextual nuances of interpersonal harm. Technological tools, including bots, operate based on predefined sets of actions, which can fall short when handling complex human interactions. In the case of ApoloBot, moderators are required to specify one offender and one victim for the apology process. However, participants noted that real-life dynamics are not always so clear-cut. Conflicts can involve multiple offenders and victims, their roles can overlap as the case escalates, or victims may remain completely unidentified. This challenge raises questions of when and how to balance the use of automation with human judgment. 

\subsubsection{Timing} Interpersonal harm often arises spontaneously and escalates unpredictably, making it difficult to determine an optimal time for tool intervention. There exists a niche window for appropriate use: a late response may allow issues to escalate, requiring more serious moderation actions, while early intervention by proactive moderators may negate the need for tools. Similarly, participants highlighted tensions regarding harm frequency: More frequent harm means more chances to utilize these tools, but it may also indicate the server's more permissive norms toward negative behaviors that can diminish successful restorative interventions. Yet too little harm~---~and correspondingly fewer opportunities to use tools like ApoloBot~---~may lead moderators to prefer manual interventions since \textit{``If it only happens on occasion, it's much easier to take care of them ourselves.'' }(P3). However, some participants saw value in having restorative tools as a safeguard, even if rarely used: \textit{``Having a bot is like a preventive measure. You have it in case something happens, that's why everyone has anti-raid even if they've never been raided.''} (P6). 

\subsection{Handling stakeholders' dropouts}
Interpersonal harm involves multiple stakeholders, and beyond their willingness to participate, it's important to consider their willingness to commit to the process throughout. In any tool that requires sustained participation, dropout will occur, and it is not yet known what the impact of partially-completed apologies might be.

\subsubsection{Victims' reasons for dropout} Victims may change their minds as the situation develops, reconsidering whether an apology would suffice to address the harm done, or they may request an apology but then be unsatisfied with the response. In these cases, moderators might need to do follow-up to understand the issue and determine what alternative steps should be taken. The worst case might be when victims receive no response at all to a request for an apology~---~as opposed to a direct negative response~---~since they may lose trust in the system and experience further emotional harm. As P12 noted:
\begin{quote}
    \textit{``Let's say the victim in their request, they were very heartfelt and genuinely wanted an apology, and fully expecting the offender to provide it. But if they can't receive a response, maybe either the offender just ran away with it or they responded in a way that the moderator deemed harmful. It's often disheartening to know that you've kind of bared your heart open to someone and they've taken advantage of it.''}
\end{quote}
\subsubsection{Offenders' reasons for dropout} Offenders unwilling to apologize may not learn from their mistakes, but even those with the intent to apologize may not always do so effectively. Participants compared ApoloBot to an appeal system, a similar framework where banned users are given chances to request unbans. They observed that it is common to receive inadequate appeals, sometimes with satisfactory ones coming after multiple attempts as offenders receive feedback, reflect, and revise their submissions. Therefore it is likely that apologies might not be at their best quality on the first try, requiring further guidance from moderators. Similarly, when offenders want to mend the relationship but their apology is rejected, they may experience significant emotional distress.

These dropout situations necessitate more thorough intervention, as the tool alone may not fully resolve the issue. More extensive follow-up actions beyond merely accepting or declining apologies may be required, or moderators may need to engage in less structured approaches in handling such cases.

\subsection{Overcoming negative perceptions of technological tools} \label{challenge-perception}
Embedding restorative justice in technical tools such as Discord bots can help initiate and facilitate communication among members, yet this kind of mediation might be perceived negatively under certain conditions. Despite human involvement in crafting the messages (apology request and response), the delivery through a bot might reduce its perceived authenticity. P15 highlighted this concern: \textit{``I think it's just how people interpret bot interactions. And we're very much used to chatbots on websites that are not useful and aren't controlled by a real person overseeing them.''} This inherent skepticism towards bots is rooted in their common stereotype that bots are impersonal and unhelpful. On Discord, this is exacerbated by prominent bot issues such as scams, phishing, and hacking, making users highly cautious when interacting with new tools regardless of their purpose. P3 described how initial negative perception can manifest into misconception, which fueled further negativity during their ApoloBot deployment:
\begin{quote}
    \textit{``Unfortunately we didn't quite well inform people about what it was. And that led to a group of people that were in this sort of echo chamber, where they were sharing misinformation about the bot due to their false understanding of it. [...] The primary one was about data collection. From my understanding, they thought that the bot was automatically collecting data about what everyone said, like an AI tool almost. And then that spread between some people and they were concerned, and be like, we don't want that. What they mentioned was just flat out wrong so we had to come in later and correct in greater detail of what data is collected, what the bot is about, and how it works.''}
\end{quote}
As P3 reflected, this incident quickly spread, leading to large-scale resistance among members, even persisting after clarification. This demonstrated how negative perceptions can create lasting barriers to tool adoption, if not addressed early and thoroughly.




\section{Discussion}



\revision{
By introducing ApoloBot as a tangible system that embeds restorative justice principles, we built on the more theoretical and hypothetical explorations of these concepts in the prior literature that influenced this work~\cite{Schoenebeck2021a, Schoenebeck2021b, Schoenebeck2023, Xiao2022, Xiao2023}. This enabled moderators to engage with these concepts in a practical manner and to envision how such tools might operate within their workflows. From this, our opportunity space uncovers the specific conditions where such tools can be applied, revealing nuances in community reception both in contexts where tools can thrive and where they may face resistance. Our findings offer a breadth of considerations that should be taken into account when developing future restorative justice tools. In this section, we expand our focus to discuss how we can translate these insights into the evaluation of future restorative justice tools, and we explore the broader implications for their potential designs.
}

\subsection{\revision{Assessing The Expected Outcomes of} Restorative Justice Tools}

\revision{
In contrast to punitive approaches that center around content-based penalties, restorative justice seeks to repair harm by addressing the multifaceted needs of victims, offenders, and their surrounding communities~\cite{Mccold2000}. Given these core differences in process and value orientation, evaluating restorative justice tools thus requires a more holistic approach that moves beyond the standard quantitative metrics such as utilization rates or punishment frequencies to focus on how effectively these tools meet the needs they are designed to address. These needs, however, are inherently personal and context-dependent, shaped by individual circumstances and by the goals of each specific community.
Therefore, a meaningful evaluation should begin with identifying \textit{what one aims to achieve}~---~the varied outcomes shaped by the diverse stakeholders needs~---~then exploring \textit{how these outcomes can be effectively measured}, by recognizing the relevant metrics accordingly~\cite{Llewellyn2013}.
}

\subsubsection{\revision{Understanding what one aims to achieve: Stakeholder's expected outcomes}}
Our study shed some lights on how the different needs of the involved stakeholders~---~moderators, victims, and offenders~---~shape their diverse expectations for restorative tools like ApoloBot, with each group defining expected outcomes in different ways depending on their unique experiences and goals.

For \textit{moderators}, 
\revision{the tool’s objective lies not in how often it is used but in its ability to reinforce community cohesion, trust, and meaningful interactions. When conditions aligned, \revisionn{two} moderators (P7, P11) found tangible outcomes from ApoloBot in resolving conflicts and fostering better relationships among members. Without direct usage due to differing community dynamics and circumstances, however, some moderators saw the same value in its mere presence.}
For them, the bot serves as a preventive measure~---~a tool that is useful and enhances safety simply by being available, even if rarely utilized. As P6 noted, \textit{"Maybe the time is not there yet"}, but having ApoloBot ready for future cases is itself a form of \revision{preemptive utility, a safeguard that contributes to a more proactive environment.} Critically, some moderators point out that too frequent usage might signal a problem, indicating recurring harm and deteriorating community relationships. However, other moderators oppose this viewpoint, \revision{favoring alternatives such as manual intervention or fully-automated sanctions over ApoloBot when visible outcomes are not yet apparent. This preference was most notable in smaller, closer-knit servers that value personal interaction, or in larger, more ``commercial'' communities that prioritize scalability. ApoloBot occupies a unique middle ground on the moderation spectrum between ``automated'' and ``manual''~\cite{Jiang2023}, facilitating victim-offender communication by automating messages and embedding participation options within its features. Communities at either extreme may therefore take longer to build the trust needed to fully utilize the tool. This aligns with findings from other socially-engaged tools, such as Chillbot, where “Goldilocks zone”, mid-sized moderation teams are those that could most effectively integrate a proactive, nudge-based approach with critical results~\cite{Seering2024}.
}

For \textit{victims} and \textit{offenders},  \revision{the preferred outcome for} usage of ApoloBot \revision{might be} less about completing a full restorative procedure and more about having access to an appropriate diversity of options to address harm. ApoloBot gives community members the autonomy to accept or decline participation based on their willingness to engage in the restoration process. \revision{As shown in our findings, while certain cases resulted in successful resolutions, sometimes both offender and victim may drop out for various reasons. If they do so,} it remains an open question whether this should be considered \revision{a desirable outcome}, as they are empowered to choose their preferred path, or a \revision{setback}, as a full resolution was not achieved. Sometimes reconciliation might not be the end result to aim for~---~people may seek other forms of closure: \revisionn{v}ictims might prefer to punish the offender~\cite{Xiao2022, Aliyu2024}, move past the situation~\cite{Xiao2023}, or leave the community entirely~\cite{Thomas2022}; \revisionn{l}ikewise, an offender might favor a more straightforward punitive process~\cite{Xiao2023} or choose to leave~\cite{Gao2024}. In this light, "success\revision{ful outcomes}" might mean enabling participants to express and fulfill their needs even if it doesn't result in an agreement\revision{, allowing the process itself to be as fluid as the human relationships it seeks to restore. }


\subsubsection{Assessing how the outcomes can be effectively measured: Potential metrics for evaluation}
\revision{
Evaluating these diverse, nuanced outcomes is thus no simple task, as has been widely discussed in research on alternative moderation strategies~\cite{Xiao2023, Ma2023, Schoenebeck2021a}.
Recognizing the breadth of different stakeholders' goals uncovered in our findings is a crucial first step toward determining how we can effectively measure them. For example, cases where victims and offenders fail to reach consensus align with real-world restorative processes, where partial restoration still holds value~\cite{Mccold2000} and outcomes can still be assessed through personalized metrics such as stakeholder satisfaction~\cite{Vanfraechem2004}, victims' feelings of security~\cite{Strang2003}, or offenders' feelings of remorse~\cite{Rowe2002}. Online communities could adapt these measures by tracking stakeholder satisfaction and emotional responses, even for incomplete processes, to better understand the factors behind partial engagement and to identify which stakeholder needs were met or unmet. At the same time, moderators can use these insights to
assess how the tool shapes the interpersonal dynamics within their communities. Offender recidivism~\cite{Bonta2002} can also be integrated as a measure for behavioral change by using the tool's log history as part of violator profiling~\cite{Cai2021} and reflection process~\cite{Cullen2022}. These community- and stakeholder-focused metrics could help moderators fully capture how well the tool upholds their values, thereby refining its application and adjusting their usage expectations. Future work can explore these evaluation metrics to inform the deployment of restorative justice tools, ensuring that they are tailored to meet the varying needs of stakeholders and fostering more adaptive, context-sensitive interventions. 
}

\subsection{Design Implication\revision{s} for Future Restorative Justice Tools}

\revision{
While ApoloBot serves as a starting point for exploring community reactions to one implementation of a restorative approach, its focus is limited to specific aspects of restorative justice: namely, facilitating apologies. The limitations of the deployment, as discussed above, therefore point to areas where current practice may need to be refined or where complementary strategies might be considered, setting the stage for future research. Building on these observations, we propose a set of design implications to guide the development of more comprehensive and impactful future restorative justice-oriented tools.
}


\subsubsection{Enabling rich interaction among stakeholders}
As discussed in Section 7.1, restorative justice tools must focus on the process of addressing stakeholder needs, and may even need to prioritize this over reaching any single outcome. This can be achieved, in part, by fostering richer and more meaningful dialogues among involved participants. ApoloBot's current design, while straightforward, only enables simple one-to-one exchange through apology requests and responses, which might not suffice in more complex cases where participants need deeper engagement. This limitation might contribute to participant dropouts, when victims are left dissatisfied or offenders struggle to express remorse effectively. To address this challenge, future tools might incorporate features for more dynamic interactions: Offenders might, for example, revise and refine their apologies based on moderator or victim feedback, reflecting genuine remorse through multiple attempts. Moderators, in turn, could offer customizable guidance, helping both parties articulate their needs and work toward meaningful resolutions.
Furthermore, tools could extend beyond dyadic exchanges to support multi-party interactions, addressing cases with overlapping roles or multiple stakeholders.
More sophisticated platforms, such as Keeper~\cite{Hughes2020}, have demonstrated the potential of restorative justice circles with socially-enhanced features like tone-setting and structured turn-taking. However, such advancements should carefully balance the benefits of improved engagement with the potential burden of increased complexity for users.


\subsubsection{Enhancing trust towards socio-technical tools}
As illustrated by P3’s experience (detailed in section \ref{challenge-perception}), miscommunication combined with negative perceptions of automated moderation may erode trust and discourage engagement. This challenge is not unique to restorative justice tools but reflects broader concerns about automated systems, where distrust often stems from users’ lack of understanding about how these tools work, coupled with prior negative experiences that tarnish their perceptions~\cite{Lee2018, Hoffman2013, Kuo2023}. These issues are exacerbated by the novelty of restorative justice itself, which remains unfamiliar to many.
To address this, interface transparency has been shown to be a key factor in improving user acceptance~\cite{Kizilcec2016}, encouraging more informed interactions~\cite{Eslami2019}, and in some cases, driving long-term moderation outcomes~\cite{Jhaver2019}. Tool developers can create clear, accessible documentation or onboarding mechanisms that can be embedded or prominently displayed with the tool to clearly explain its functionalities and the human-centered restorative justice principles behind its design. Moreover, additional work can be done to explore how this explanation can be made more personalized and "humanized" to overcome the perceptions of insincere bot-driven interactions, especially when dealing with emotional matters like interpersonal harm.



\subsubsection{Resource sharing and education}
The opportunity space explored in our findings highlighted groups of moderators who showed interest in ApoloBot's approach but were not yet positioned to fully utilize it. For instance, very proactive moderation teams had interest in re-engaging in restorative conversations but lacked the mediation skills necessary to do so. On the other end of the spectrum, moderators in communities with much more lax behavioral norms might perceive the tool as high-risk due to the potential for abuse, or for less sincere engagement with the tool.
Nevertheless, these outcomes are not necessarily inevitable~---~community moderation is an evolving practice that grows through collective learning within~\cite{Cullen2022} and across~\cite{Hwang2024, Uttarapong2024} communities. Reflective practices and social learning have been shown to close knowledge gaps and reshape perceptions of tool adoption, transforming it from a burdensome transition into an engaging and even enjoyable experience~\cite{Hwang2024}. By fostering this collaborative process, a gradual shift toward deployment of restorative justice-based tools can be made more accessible when communities can share and refine their technological frame through an emergent social ecosystem. Technological platforms or frameworks enabling communities to document and share their encountered use cases can be valuable in providing inspiration and practical insights for communities to address gaps in values, functionalities and possible scenarios for restorative justice, allowing communities to adapt and align tools within their own contexts accordingly.

In addition, education plays a vital role not only for moderators but also for community members~\cite{West2018}, many of whom may lack prior exposure to restorative justice processes. Future work can further explore how educational initiatives can be informed through platform or tool design, such as via explanations, guided interactions, or prompts to provide users with actionable insights that can ease their navigation through restorative interactions. For example, in the case of ApoloBot, providing guidance on how to write a "reasonable" apology can help offenders construct meaningful responses that can better meet victims' needs. These educational elements can be further tailored to specific community values, as informed by collective learning and experimentation, to create an environment where restorative outcomes are both practical and well-aligned with the local cultural dynamics.

\subsubsection{Beyond harm resolution}
While ApoloBot's focus on harm resolution offers key insights into the core aspects of restorative justice implementation, it also carries limitations including aspects outside its current scope, such as timing before intervention or dropout follow-ups. On this account, future restorative justice-oriented tools can expand their capabilities to encompass a broader spectrum of the moderation process, addressing not only the harm by itself but also the stages \textit{pre-harm} and \textit{post-harm}.

For \textit{pre-harm interventions}, a proactive identification approach can play a crucial role in determining the right moment for restorative actions. In our study design, the harm identification process is fully manual, posing challenges related to moderator workload and timing of interventions~---~acting too early risks unnecessary interference, while acting too late can exacerbate harm. Recent advances in human-AI moderation systems have shown promise in identifying potential at-risk interactions that fall within this optimal window for restorative intervention. By analyzing relevant contextual signals such as conversation metrics~\cite{Choi2023, Schluger2022}, user histories~\cite{Im2020}, and prior moderation decisions~\cite{Chandrasekharan2019}, technologies can guide moderators’ attention to potential harm cases more efficiently, especially in large and high-traffic communities. This opens up opportunities for designing detection mechanisms specific to community dynamics that can be embedded as a more context-aware initiation of restorative workflows. 

For \textit{post-harm interventions}, a support network might provide additional avenues for engagement and healing, especially in cases where dropout occurs. Victim-support groups, as a form of restorative justice~\cite{Dignan2004}, have been previously discussed and implemented in the form of online systems designed for emotional support, advice sharing and empowerment~\cite{Dimond2013, Blackwell2017}. At the same time, it could also be valuable to explore similar consultation mechanisms for offenders, particularly those who genuinely wish to or have the potential to engage in the restorative process. Capturing the effects of these interventions and monitoring behavioral trajectories \textit{post-harm} could also provide valuable context to strengthen and tailor \textit{pre-harm} strategies. For instance, understanding patterns of accountability, remorse, or recurrence among offenders could inform predictive tools and guide proactive measures to mitigate harm.


Overall, restorative justice-oriented tools offer a broad design space, encompassing a spectrum of possibilities. They might involve \textit{embedded features} that enhance existing tools with functionalities like transparency or user feedback. At the next level, they could include dedicated \textit{bots} designed for specific restorative tasks, such as ApoloBot’s focus on facilitating apologies. For more comprehensive needs, \textit{integrated systems} could combine tools—such as pairing a bot with harm identification mechanisms—to offer greater functionality. At the highest level, \textit{platforms} like HeartMob, Keeper or knowledge-sharing spaces can provide fully realized ecosystems for restorative justice practices. By accommodating varying degrees of complexity, these tools can be customized to fit existing infrastructures and community-specific needs. This expanded vision reimagines restorative justice tools not only as solutions for conflict but also as catalysts for more proactive, inclusive, and resilient community governance.
\section{Conclusion}

\revision{While restorative justice has demonstrated its potential in offline contexts, its implementation in online spaces remains a relatively new challenge. Motivated by this gap, our study introduces} ApoloBot\revision{, a Discord bot that integrates restorative principles through apology facilitation,} as a method for exploring the design space of restorative justice tools. Through interviews and a deployment with active Discord moderators, we identified key opportunities and challenges associated with these tools, considering various factors spanning the community values, moderation practices, and case scenarios that influence their effectiveness. We also \revision{expanded our focus to discuss} the evaluation for \revision{restorative outcomes,} along with \revision{broader} implications for future initiatives. Our work sets forth a foundation for the future design of online restorative justice tools, offering insights into their \revision{viability} and areas for further development.


\begin{acks}
This work was supported by the National Research Foundation of Korea (NRF) grant funded by the Korea government (MSIT) (RS-2024-00348993), as well as by a KAIST Undergraduate Research Program grant. EPFL CDH also provided financial support for the first author to attend the conference and present the work. We thank the community moderators for their active engagement and thoughtful feedback throughout the study. We also thank all members of CSTL for their constructive discussions and invaluable support.
\end{acks}

\bibliographystyle{ACM-Reference-Format}
\bibliography{main}

\appendix
\section{Questions for Survey/Interviews}
\subsection{Survey Questions for Recruitment}
\textbf{\textit{General Question}}
\begin{enumerate}
    \item How long have you been moderating on Discord community/communities? [Multiple Choice]
    \begin{itemize}
        \item Less than 6 months
        \item 6 months to 1 year
        \item 1 to 2 years
        \item 2 to 3 years
        \item 3 to 5 years
        \item More than 5 years
    \end{itemize}
\end{enumerate}

\textit{\textbf{Interpersonal Harm on Online Communities:} Interpersonal harm here means one-to-one conflicts or misbehaviors that involve distinct offender and victim. For example, if someone directs hurtful language, embarrassment or threat towards another individual, whether intentionally or unintentionally, it qualifies as interpersonal harm. However, issues like spamming or hate speech towards general sub- groups may not fall under this category.}

\begin{enumerate}
    \item Have you ever encountered situations involving interpersonal harm within your Discord community? If yes, please briefly describe one such situation, and how it was handled. [Long Answer]
    \item How often does this kind of case happen in your Discord community? [Long Answer]
\end{enumerate}

\textit{\textbf{Moderation Experience}: For this section, please answer the following questions about the Discord server(s) that you currently moderate where you most frequently encounter interpersonal harm, as mentioned in the previous section.}
\begin{enumerate}
    \item What are the specific types of these server(s)? (e.g Overwatch game, BTS fandom, Coding tutorials, etc) [Short Answer]

    \item What is the size of the server(s) you've just mentioned? [Checkboxes]
    \begin{itemize}
        \item Less than 100 members
        \item 100 to 500 members
        \item 501 to 1,000 members
        \item 1,001 to 5,000 members
        \item 5,001 to 10,000 members
        \item More than 10,000 members
    \end{itemize}

    \item What tools do you regularly use for moderating these server(s)? [Long Answer]

    \item What are the primary roles or tasks you perform as a moderator in these server(s)? [Long Answer]

    \item (Optional) Provide a link to the server(s) or any relevant webpage that contains information about the community (e.g link to Discord server, subreddit, forum, etc) [Short Answer]
\end{enumerate}

\subsection{Interview Questions for Phase 1}
The interview sessions follow a semi-structured format, where the researcher follows a core set of questions while allowing flexibly to adjust the flow and explore follow-up topics. Below is the initial set of questions:

\textit{\textbf{Ice Breaking and Introduction}}
\begin{enumerate}
    \item How did you become a moderator in your community? 
    \item From when you started until now, how has your journey as a community moderator been?
    \item What tasks do you do as a moderator?
    \item Which tools or methods have you found most effective in making your moderation tasks easier and more efficient?
\end{enumerate}

\textit{\textbf{Project Presentation and Discussion on Interpersonal Harm}}

\textit{Researcher explain about the concept of Interpersonal Harm on Online Communities.
The following questions refer to the examples of interpersonal harm that participant gives in the screening survey.}
\begin{enumerate}
    \item Can you specify in more details about the case you mentioned?
    \item What was your decision rationale in this case?
    \item What was the impact of this case on the people involved?
    \item Was there also any impact that this situation had on the community as a whole?
    \item Are there any defined procedure for dealing with this type of harm?
    \item Can you also specify a different case of interpersonal harm that has happened?
\end{enumerate}

\textit{\textbf{Research Work and Discussion on ApoloBot}}

\textit{Researcher explains about the concept of Restorative Justice and introduces ApoloBot.}

\begin{enumerate}
\item Do you think that ApoloBot would be appropriate (or suitable) in your community, given the culture or its overall dynamics?
\item Linking back to the case we’ve mentioned earlier, would ApoloBot be useful in this specific scenario?
\item What kinds of cases it might be useful?
\item What kinds of cases it might not be as useful?
\item Overall, what are your perceived pros and cons of using ApoloBot in your online communities?
\item How might ApoloBot be incorporated into the existing workflow?
\item Would there be any challenges for this integration?
\item How willing and able do you feel to facilitate this apology process via ApoloBot as a moderator?
\end{enumerate}

\textbf{\textit{Free-style discussion on general restorative justice tools}}

\subsection{Interview Questions for Phase 3}
Similarly, Phase 3 interview sessions were semi-structured but allowed for even greater flexibility as participant had varied experience with ApoloBot. We based on the following sets of questions, incorporating follow-ups and other remarks while extending the discussion to include restorative justice tools more broadly.

\textbf{\textit{General Questions}}
\begin{enumerate}
    \item How was your general experience with ApoloBot?
    \item Does this match your initial expectation when you start using it?
\end{enumerate}

\textbf{\textit{Reflection - On community and moderation practices}}

\textit{If ApoloBot was used:}
\begin{enumerate}
    \item Does using the bot change the practice where you moderate harm cases in your community?
    \item How did you utilize this tool in combined with the existing moderation system?
    \item Were there any perceived impact for the community? 
    \item Will the bot still be valuable if it continues to be used in the community?
    \item Was there any reaction or perception from other moderators in the team, observing how the tool is built and used?
    \item Will they be willing to use ApoloBot if they also have the chance to?
    \item Other than the perceived benefits, were there also any challenges or obstacles you encountered when using ApoloBot?
\end{enumerate}

\textit{If ApoloBot was not used:}

\begin{enumerate}
    \item What were the main reasons for not using ApoloBot in your community?
    \item Did you encounter some challenges when using it?
    \item Was there any reaction/perception from other moderators in the team, observing how the tool is built?
    \item What kind of harm happened recently in the community during the testing period?
    \item And why was ApoloBot not used for these cases?
\end{enumerate}

\textbf{\textit{Reflection - On case scenarios}}

\textit{If ApoloBot was used:}

\begin{enumerate}
    \item Can you walk me through the contexts of the specific cases ApoloBot was used?
    \item What went well?
    \item What didn’t? 
    \item In cases where the bot works, how did the involving members react to having ApoloBot facilitating their conversations?
    \item Are there any interesting insights you encountered or lessons learned when resolving cases with ApoloBot?
\end{enumerate}

\textit{If ApoloBot was not used:}
\begin{enumerate}
    \item What type of other cases that happened during the testing period? 
    \item Why ApoloBot was not used in these cases?
    \item Will there still be value if ApoloBot was not used frequently as such?
\end{enumerate}

\textit{For all:}

\begin{enumerate}
    \item Were there cases where:
    \begin{itemize}
        \item You tried but it didn’t work?
        \item You thought of using but ended up didn’t
    \end{itemize}
\end{enumerate}

\textbf{\textit{Free-style discussion on general restorative justice tools}}


\end{document}